\documentclass[preprint,aps,showpacs,nofootinbib]{revtex4}

\usepackage{graphicx}
\usepackage{epsfig,amssymb,amsmath,color}
\usepackage{epstopdf}

\def\f{\phi}

\newcommand{\beq}{\begin{equation}}
\newcommand{\eeq}{\end{equation}}
\newcommand{\bea}{\begin{eqnarray}}
\newcommand{\eea}{\end{eqnarray}}
\newcommand{\bear}{\begin{array}}
\newcommand {\eear}{\end{array}}
\newcommand{\bef}{\begin{figure}}
\newcommand {\eef}{\end{figure}}
\newcommand{\bec}{\begin{center}}
\newcommand {\eec}{\end{center}}

\newcommand{\la}{\left\langle}
\newcommand{\ra}{\right\rangle}

\def\EQ#1{Eq.~(\ref{#1})}

\def\REF#1{(\ref{#1})}
\def\GEV#1{10^{#1}{\rm\,GeV}}

\def\lrfp#1#2#3{ \left(\frac{#1}{#2} \right)^{#3}}

\begin{document}
\draft
\tighten
\preprint{KEK-TH-1717, TU-962, 
IPMU14-0068
}
\title{\large \bf
Multi-Natural Inflation in Supergravity and BICEP2
}
\author{
Michael Czerny\,$^{a,\ast}$\footnote[0]{$^\ast$ email: mczerny@tuhep.phys.tohoku.ac.jp},
Tetsutaro Higaki\,$^{b,\star}$\footnote[0]{$^\star$ email: thigaki@post.kek.jp},    
Fuminobu Takahashi\,$^{a,\,c\,\dagger} $\footnote[0]{$^\dagger$ email: fumi@tuhep.phys.tohoku.ac.jp}
    }
\affiliation{
    $^a$ Department of Physics, Tohoku University, Sendai 980-8578, Japan \\
    $^b$ Theory Center, KEK, 1-1 Oho, Tsukuba, Ibaraki 305-0801, Japan \\
    $^c$ Kavli IPMU, TODIAS, University of Tokyo, Kashiwa 277-8583, Japan
    }

\vspace{2cm}

\begin{abstract}
We revisit the recently proposed multi-natural inflation and its realization in supergravity
in light of the BICEP2 results. Multi-natural inflation is a single-field inflation model
where the inflaton potential consists of multiple sinusoidal functions, and it is known that
a sizable running spectral index can be generated, which relaxes the tension between
the BICEP2 and the Planck results.  In this paper we show that multi-natural inflation can
accommodate a wide range of values of $(n_s, r)$, including the spectral index close to or even above unity.
This will be be favored if the tension is resolved by other sources such as dark radiation, hot dark matter, 
or non-zero neutrino mass. We also discuss the implications for the implementation 
in string theory.
\end{abstract}
\maketitle


\section{Introduction}

The BICEP2 experiment detected the primordial B-mode polarization~\cite{Ade:2014xna}, which could originate
 from the gravitational waves generated during inflation~\cite{Guth:1980zm,Linde:1981mu}.
The tensor-to-scalar ratio $r$ and the Hubble parameter during inflation  $H_{\rm inf}$ suggested by the
BICEP2 are given by
\bea
\label{B-Hr}
H_{\rm inf} &\simeq& 1.0 \times \GEV{14} \lrfp{r}{0.16}{\frac{1}{2}},\\
r &=& 0.20^{+0.07}_{-0.05} ~~(68\%{\rm CL}).
\label{B-r}
\eea
The preferred range of $r$ is modified to $r = 0.16^{+0.06}_{-0.05}$ after subtracting the best available estimate for foreground dust.

Taken at face value, the BICEP2 results strongly suggest large-field inflation such as chaotic inflation~\cite{Linde:1983gd}.
 For various large-field inflation models and their concrete realization in the standard model as well as supergravity and superstring theory,
see e.g.~\cite{Freese:1990ni,Murayama:1992ua,Kawasaki:2000yn,Dimopoulos:2005ac,Kallosh:2007ig,Silverstein:2008sg,McAllister:2008hb,Kaloper:2008fb,Takahashi:2010ky,
Nakayama:2010kt,Nakayama:2010sk,Harigaya:2012pg,Croon:2013ana,Nakayama:2013jka,
Nakayama:2013nya,Cicoli:2014sva,Czerny:2014wza,Czerny:2014xja,Nakayama:2014-HCI}. It is worth noting, however,  that the BICEP2 results are in tension with the {\it Planck} + WP + highL data. Explicitly, there is a tension on the relative size of scalar density perturbations on large and small scales. This tension suggests another
extension of the Lambda CDM model such as a running spectral index, dark radiation, hot dark matter (HDM), 
 non-zero neutrino mass~\cite{Ade:2014xna,Giusarma:2014zza}, 
 anti-correlation between tensor and scalar modes~\cite{Contaldi:2014zua} or between tensor 
 and isocurvature modes~\cite{Kawasaki:2014lqa}, and a sharp cut-off in the scalar 
 modes~\cite{Miranda:2014wga}.\footnote{See  e.g. Refs.~\cite{Feng:2003zua,Kawasaki:2003dd}
 for double inflation models to realize a sharp cut-off in the curvature perturbations.}
If it is due to the running spectral index, it would provide us with invaluable information on the inflation sector.\footnote{
The implications of the other possibilities are also significant. The HDM candidates include axions or sterile neutrinos.
The QCD axion can explain both hot and cold dark matter~\cite{Jeong:2013oza}.
If both sterile neutrino HDM and dark radiation are present, there will be three coincidence problems of
baryon-dark matter, neutrino-sterile neutrino, and photon-dark radiation densities. 
This may suggest the dark parallel world~\cite{Higaki:2013vuv}.
} 

The BICEP2 data implies that the inflaton field excursion exceeds the 
Planck scale, $M_{pl} \simeq 2.4 \times \GEV{18}$~\cite{Lyth:1996im} . In order to have a sensible inflation model with super-Planckian field values, one has to control
the inflaton potential over a broad field range.  One way to accomplish this is to introduce a
shift symmetry, under which the inflaton $\phi$ transforms as
\bea
\phi &\to& \phi +C,
\eea
where $\phi$ is the inflaton and $C$ is a real transformation parameter. 
As we shall see later, one of the plausible candidates for the inflaton with the above property is an axionic component of a moduli field.
In fact, there appear many moduli fields  through compactifications in string theory, and one of them could
be the inflaton. 

The shift symmetry needs to be explicitly broken by some sources to generate the inflaton potential,
and one plausible functional form is the sinusoidal function. In this case, the continuous shift symmetry 
is broken down to a discrete one.  If a single sinusoidal function dominates the inflaton potential, it is the 
natural inflation~\cite{Freese:1990ni}. On the other hand, if there are many sources for the explicit breaking, 
it is the multi-natural inflation proposed in Ref.~\cite{Czerny:2014wza}, and its realization in supergravity 
was given in Ref.~\cite{Czerny:2014xja}. In multi-natural inflation, 
the inflaton potential consists of multiple sinusoidal functions with different height and periodicity:
\bea
\label{Vmulti}
V_{\rm multi-natural} &=& \sum_i \Lambda_i^4 \cos\left(\frac{\phi}{f_i} + \theta_i \right) + {\rm const.},
\eea
where the last constant term is such that the inflaton potential becomes zero at the potential minimum.
Interestingly, only with two sinusoidal functions,  a wide range of values of $(n_s, r)$ can be realized.\footnote{
Even small-field inflation is possible with two sinusoidal functions.
See Ref.~\cite{Takahashi:2013tj} for curvatons with multiple sinusoidal functions.
}

We would like to emphasize the difference of the multi-natural inflation from $N$-flation~\cite{Dimopoulos:2005ac}.\footnote{
The realization of the $N$-flation in Large Volume Scenario~\cite{Balasubramanian:2005zx} was given
in Ref.~\cite{Cicoli:2014sva}.}
In the case of $N$-flation, there are many $(\gtrsim 100)$ axions, each of which has 
a sinusoidal potential. As a result, the high-scale inflation can be realized with sub-Planckian
decay constants. In contrast, the multi-natural inflation is a single-field inflation, whose potential
consists of multiple sinusoidal functions as in (\ref{Vmulti}), and the inflaton dynamics is
much simpler.

Recently, it was shown in Ref.~\cite{Czerny:2014wua} that
multi-natural inflation has a built-in feature for generating a sizable running spectral index. This is because,
if the inflaton potential has small modulations, a sizable running spectral index can be generated without
spoiling the overall inflaton dynamics, as pointed out in Ref.~\cite{Kobayashi:2010pz}. 
The running spectral index, if confirmed by other experiments such as the 21cm observations~\cite{Shimabukuro:2014ava}, would tell us
about rich structures of the inflaton potential. On the other hand, it is also possible that the tension between the
BICEP2 and the Planck is due to some other sources.  Even in this case, multi-natural inflation remains 
one of the viable large-field inflation models, which is one of the main targets of this paper. 

In this paper we revisit multi-natural inflation and its realization in supergravity in light of the recent BICEP2
results, and show that a wide range of values of $(n_s, r)$ can be realized. In particular, the spectral index
close to or above unity with a large $r$ satisfying \EQ{B-r} is possible. Such large $n_s \sim 1$ will be favored
if the tension between the BICEP2 and the Planck is solved by other sources such as dark radiation,  HDM, 
or non-zero neutrino mass~\cite{Giusarma:2014zza}. We will also discuss the implications for 
the implementation of multi-natural inflation in string theory.

\section{Multi-natural inflation in supergravity}
\label{SecSUGRA}
\subsection{Setup and saxion stabilization in the case of one axion}

We shall review a realization of multi-natural inflation  in which the  axion of a modulus field plays the role
of the inflaton within the framework of supergravity~\cite{Czerny:2014xja}. 
We introduce the following effective theory with an axion chiral superfield $\Phi$ for simplicity,\footnote{
The natural inflation in supergravity is given in Ref.~\cite{Kallosh:2007ig}.
}
\bea
\label{K1}
K &=& 
\frac{f^2}{2}(\Phi + \Phi^{\dag})^2 , \\
W &=& W_0 + Ae^{-a\Phi} + Be^{-b\Phi},
\label{W1}
\eea
where  $a>0$, $b>0$, $a \ne  b$ and the following relations are assumed:
\beq
f \lesssim 1 , ~~~ |B| \lesssim |A| \ll |W_0| < 1. 
\label{condp1}
\eeq
Here and in what follows we adopt Planck units where the reduced Planck mass $M_{pl}$ is set to be unity. 
The scalar potential is given by
\bea
\label{sugra}
V= e^{K}[K^{i\bar{j}}(D_iW) (\overline{D_jW})-3|W|^2],
\eea
with $D_i W = (\partial_i K) W + \partial_i W$.
For later convenience let us express $\Phi$ as
\bea
\Phi = \sigma + i \varphi,
\eea
where we refer to $\sigma$ and $\varphi$ as the saxion and the axion, respectively. 
The kinetic term for the saxion and the axion is given by
\beq
{\cal L}_{\rm kin} \;=\; K_{\Phi {\bar \Phi}} \partial \Phi^\dag \partial \Phi =  f^2(\partial \sigma)^2 + f^2(\partial \varphi)^2.
\eeq
The canonically normalized saxion and axion  fields are $\sqrt{2} f \sigma$ and $ \sqrt{2} f \varphi$, 
respectively.

The K\"ahler potential (\ref{K1}) respects the shift symmetry of the axion,
\bea
\varphi \to \varphi+ {\rm const.},
\eea  
which is explicitly broken by the two exponential terms in the superpotential. 
In Eq.~(\ref{condp1}), we have assumed that the breaking scale of the shift symmetry is so low 
that the axion is much lighter than the saxion. Thus, the inflation model can be effectively described 
by the single-field inflation with the axion being identified with the inflaton. 

To see how the saxion is stabilized, let us first focus on the case of $A=B=0$ for simplicity. 
The saxion is then stabilized  via the equation
\bea
\partial_\Phi K = 2 f^2  \sigma = 0.
\eea
This is because the saxion potential is given by
\bea
V &= & e^{2f^2 \sigma^2}\bigg(4 f^2 \sigma^2 - 3  \bigg) |W_0|^2 + \Delta V \\
& \simeq &   2 f^2  |W_0|^2 \sigma^2 + \cdots,
\eea
where the potential is expanded around the origin in the second equality,
%
%
and we add a sequestered SUSY-breaking potential $\Delta V$  to realize the Minkowski vacuum\footnote{
Alternatively, we can use the F-term up-lifting potential~\cite{Lebedev:2006qq,Kitano:2006wz} to break SUSY with a
vanishingly small cosmological constant in the present vacuum. 
The heavy SUSY breaking fields can be integrated out during inflation. 
See \cite{Kachru:2003aw,Choi:2005ge,Dudas:2006gr} for the related topics.
}:
\bea
\Delta V &=& 3 e^{2K/3} |W_0|^2  
\simeq \bigg(3 + 4f^2  \sigma^2 + \cdots\bigg)|W_0|^2.
\eea 
The saxion mass is given by $m_{\sigma} \simeq \sqrt{2} |W_0|$, and is therefore decoupled from the inflaton dynamics if $H_{\rm inf} < W_0$, where, as we shall see later,
$H_{\rm inf} \sim \sqrt{AW_0}$ in our scenario. 

The saxion can be similarly stabilized for a more general K\"ahler potential via the same equation of 
$\partial_\Phi K = 0$;
see Refs.~\cite{Conlon:2006tq,Choi:2006za,Higaki:2011me} for detailed discussions on the saxion stabilization.
In general, the saxion mass is considered to be on the order of the gravitino mass.

Even if we turn on the exponential terms $A, B \ne 0$, the saxion stabilization discussed above is not significantly changed and the saxion vacuum is located near the origin, as long as condition \REF{condp1} is satisfied.  
To be explicit, the above analysis remains valid if the axion
mass is  lighter than that of the saxion mass by a factor of ten~\cite{Czerny:2014xja}.

\subsection{Multi-natural inflation with two sinusoidal functions}
Once the saxion is stabilized with a heavy mass, we can focus on the axion dynamics. 
First, let us derive the axion potential.
We can set $W_0$ and $A$ real and positive using $U(1)_R$ symmetry and an appropriate shift of the axion field, 
while $B$  is complex in general. To explicitly show the complex phase,  we replace $B$ with $B e^{- i \theta}$, 
where $B$ is a real and positive constant, and $\theta$ denotes the relative phase between the two shift-symmetry breaking 
terms. 
Substituing $\la \sigma \ra \simeq 0$, the axion potential is obtained as
\bea
\nonumber
V_{\rm axion}(\phi) &\simeq&
6 A  W_0 \left[ 1 - \cos\left(\frac{\phi}{f_1}\right)\right] 
+ 6 B  W_0 \left[1- \cos\left(\frac{\phi}{f_2} + \theta \right)\right]  \\
&& - 2 A B \left( \frac{2}{f_1 f_2} -3\right) \left[
1- \cos\bigg[ \left( \frac{1}{f_1} -\frac{1}{f_2}  \right)\phi -\theta  \bigg] 
\right] + {\rm const.},
\label{pot}
\eea
where $\phi \equiv \sqrt{2} f\varphi$ is the canonically normalized axion field, and the 
two decay constants $f_1$ and $f_2$ are defined by
\bea
f_1 \equiv \frac{\sqrt{2}f}{a},~~~f_2 \equiv \frac{\sqrt{2}f}{b},
\eea
with $f_1 \ne f_2$.
The vacuum energy  is set to be zero in the true vacuum by adjusting the last constant term,
which depends on $\theta$; it vanishes for $\theta =0$.

When we impose relation (\ref{condp1}) to realize the mass hierarchy between the axion 
and the saxion,  the third term in \EQ{pot} becomes irrelevant; 
the first two terms are equivalent to the inflaton potential
for the multi-natural inflation with two sinusoidal functions studied in Ref.~\cite{Czerny:2014wza}. 

The multi-natural inflation with two sinusoidal functions can accommodate a large running spectral
index without affecting the overall inflation dynamics, provided there is a mild hierarchy in the two decay 
constants~\cite{Czerny:2014wua}, and the second sinusoidal function acts as small modulations to the
first one. To this end, the following (mild) hierarchy is required:
\bea
&& B = {\cal O}(10^{-3}) \,A,\\
&& f_2 = {\cal O}(10^{-2})\,  f_1.
\eea
In this single axion multiplet model, the latter can be realized 
when one takes $a = 2\pi/n_1$ and $b=2\pi/n_2$ with $n_1={\cal O}(100)$ and $n_2={\cal O}(1)$.

The inflation scale is then given by $H_{\rm inf} \sim \sqrt{A W_0} \ll W_0$. 
Thus, the saxion mass is in general heavier
than the Hubble parameter during inflation, justifying our previous assumption that the saxion is decoupled from the inflaton dynamics. This is a nice point against the moduli destabilization
during inflation \cite{Kallosh:2004yh}. 
Also, detection of the tensor mode by BICEP2 \REF{B-Hr} determines the typical scale
of the exponential terms as
\bea
AW_0  &=& {\cal O}( 10^{-9})
\eea
for $f_1 = {\cal O}(10)$.

\subsection{A case of two axions for a large decay constant}

We now study another realization of a large decay constant in the context of
multi-natural inflation, based on  the idea given in Ref. \cite{Kim:2004rp}.
The point is that, if there are heavy and light axions,  the effectively large 
decay constant is realized by accidental cancellation between the two axion 
potentials.

We consider the effective theory with two axions given by
\bea
\nonumber
K &=& 
\frac{f^2}{2}(\Phi_1 + \Phi_1^{\dag})^2 + \frac{f^2}{2}(\Phi_2 + \Phi_2^{\dag})^2, \\
W &=& W_0 + Ce^{-c (\Phi_1 + \Phi_2)} + Ae^{-a (\Phi_1 + (1 + \Delta_1 ) \Phi_2)} + Be^{-b (\Phi_1 + (1 + \Delta_2 ) \Phi_2)-i\theta},
\label{W2}
\eea
with $B \lesssim A \ll C \ll W_0 < 1$.
For simplicity, we use the same decay constant $f \lesssim 1$ for $\Phi_1$ and $\Phi_2$ and 
all the parameters are set real and positive and the CP phase between exponentials is expressed by $\theta$ as in the previous case.
The point is that one combination of the axions is stabilized with a heavy mass 
mainly by the first exponential term, whereas the other combination remains light and receives the 
potential from the second and third exponential terms. For a certain combination of the heavy and light
axions appearing in the exponential terms, the effective decay constant for the lighter one can be
larger than the Planck scale~\cite{Kim:2004rp}.

To be explicit, let us define the heavier axion $\sqrt{2}\xi \equiv \phi_1 + \phi_2 $ and the lighter axion 
$\sqrt{2}\phi \equiv -\phi_1 + \phi_2$, where $\phi_{1,2}$ are the canonically normalized axions of 
imaginary component of $\Phi_{1,2}$ defined by $\phi_i = \sqrt{2} f\, {\rm Im}[\Phi_i]$ with $i=1,2$.
The scalar potential for axions can be approximately given by
\bea
\nonumber
V_{\rm axion}  (\phi, \xi)
&\approx&
-6 W_0 A \cos\bigg[\frac{(2+\Delta_1)\xi + \Delta_1 \phi}{2F_1}\bigg]
-6 W_0 B \cos\bigg[\frac{(2+\Delta_2)\xi + \Delta_2 \phi}{2 F_2} + \theta \bigg] \\
&& -6 W_0 C \cos\bigg[\frac{\xi}{F_0}\bigg] + {\rm const.} 
,
\eea
where the decay constants are defined by $F_0 = f/c,~F_1 = f/a,$ and $F_2 = f/b$,
and we have neglected the cross term between shift-symmetry breaking terms.
We also dropped  the dependence on the saxions because they will be stabilized near the origin
via $\partial_{\Phi_1}K \simeq \partial_{\Phi_2}K \simeq 0$ after the up-lifting potential is added. 
For $A \ll C/F_0^2$, $\xi$ is stabilized at the origin during the inflation driven by $\phi$. Hence, 
the effective scalar potential for the lighter axion below the mass scale of $\xi$ reads
\bea
V_{\rm axion}^{({\rm eff})} (\phi)
&\approx&
-6 W_0 A \cos\bigg[\frac{ \phi}{f_1}\bigg]
-6 W_0 B \cos\bigg[\frac{ \phi}{f_2} + \theta \bigg] + {\rm const.} ,
\eea
where
\bea
f_1 \equiv 
 \frac{2}{a\Delta_1}f
,~~~f_2 \equiv 
 \frac{2}{b\Delta_2}f .
\eea
Thus, one can realize the super-Planckian decay constants $f_1 = {\cal O}(10)$
for e.g. $a=2\pi/n_1$, $n_1={\cal O}(10)$,  $\Delta_1 ={\cal O}(0.1)$ and $f = {\cal O}(0.1)$ without severe fine-tuning of the parameters, 
and  similarly for $f_2$. The mild hierarchy $f_2 = {\cal O}(10^{-2})  f_1$ required for generating the large running spectral index 
can be realized if we take $b=2\pi/n_2$ with $n_2={\cal O}(1)$, $\Delta_2 ={\cal O}(1)$.
%
%
%
%

\subsection{Toward a UV completion}

Here we discuss a  possible UV completion of the multi-natural inflation based on string-theoretic supergravity.
In the viewpoint of the UV completion, the multi-natural inflation is understood to be 
a remnant of the moduli stabilization of extra dimension 
with various non-perturbative dynamics involving sub-leading effects. 
The successful stabilization is required for a sensible string theory in four dimensions.

Toward the UV completion, we may consider
\bea
K &=& -2\log(t_0^{3/2} - t_1^{3/2}-t_2^{3/2} - t_3^{3/2} -t_4^{3/2}), \\
\nonumber
W &=& W_0 - D e^{-\frac{2\pi}{N}T_0} - E e^{-\frac{2\pi}{M}(T_1+T_2)} - F e^{-\frac{2\pi}{L}(T_3+T_4)}  \\
&&+ C e^{-\frac{2\pi}{n_0} (T_2 + T_4)}
+ A e^{-\frac{2\pi}{n_1}T_2 }+ B e^{-\frac{2\pi}{n_2}(T_2+ 2 T_4)-i\theta},
\label{siW}
\eea
where $t_i = (T_i +T_i^{\dag}) ~{\rm for}~i=0,1,2,3,4$, and $T_i$ are complex K\"ahler moduli on a Calabi-Yau space, 
and
$W_0,~A,~B,~C,~D,~E$ and $F$ are determined by
the vacuum expectation values (VEVs) of heavy moduli via flux compactifications~\cite{Grana:2005jc,Blumenhagen:2006ci}. 
(See \cite{Giryavets:2003vd} for  realization of a small $W_0$.)
The non-perturbative terms in the superpotential are assumed to be generated by gaugino condensations in a 
pure $SU(N)\times SU(M)\times SU(L) \times SU(n_0) \times SU(n_1) \times SU(n_2)$ gauge theory living on D-branes. 
The holomorphic gauge couplings are given by the moduli 
$T_0$, $T_1+T_2$, $T_3+T_4$, $T_2+T_4$, $T_2 $ and $T_2 + 2T_4$ respectively. 
We assume that those parameters satisfy
\bea
&& A,  B, C, D, E, F = {\cal O}(1),~~~W_0 \ll 1.
\eea
All the  above parameters are set real and positive using
 U(1)$_R$ symmetry and an appropriate shift of the imaginary components of the moduli fields,
  while $\theta$ represents the relative phase among the exponentials. 
We also assume the relation between the parameters in the exponents
\bea
&& n_0 \gtrsim n_1 \gtrsim n_2 ={\cal O}(1-10), \\
&& L \sim M \sim N > 2n_0 .
\label{n1MN}
\eea 
The mild hierarchy between them makes $T_0$, $T_1+T_2$ and $T_3 +T_4$ stabilized in a
supersymmetric manner by the first three exponentials \cite{Kachru:2003aw,Choi:2006za,Czerny:2014xja}:
\bea
&&\frac{2\pi}{N}\langle T_0 \rangle \sim \frac{2\pi}{M}\langle T_1+T_2 \rangle \sim
\frac{2\pi}{L}\langle T_3 + T_4 \rangle \sim \log (1/W_0),
\eea
and the last two terms in the superpotential are responsible to the multi-natural inflation thanks to such a hierarchy.
On the other hand $\Phi_1 = -T_1 + T_2$ and $\Phi_2 = -T_3 +T_4$ remain relatively
light; these combination becomes two axion supermultiplets in the previous subsection,
whose effective action is similar to Eq.(\ref{W2}) with $a=\pi/n_1$, $b=\pi/n_2$, $c=\pi/n_0$, $\Delta_1 = -1$ and $\Delta_2 = 1$.
They will be stabilized near the origin through the equation of
$\partial_{\Phi_1} K \simeq  \partial_{\Phi_2} K \simeq 0$, i.e., ${\rm Re}[\Phi_1] \simeq {\rm Re}[\Phi_2] \simeq 0$
after the sequestered SUSY-breaking term is added:
\bea
V = V_{\rm moduli} + V_{\rm up},~~~{\rm where}~~
V_{\rm up} = \hat{\epsilon}\, e^{2K/3};~~~\hat{\epsilon} ={\cal O}(W_0^2).
\eea
The decay constant $f$ is determined by the Calabi-Yau volume constituted by
three heavier moduli VEVs, and expected to be of order string scale;
$f \sim 1/((t_1+t_2)^{1/4}{\cal V}^{1/2}) \sim 1/((t_3+t_4)^{1/4}{\cal V}^{1/2})  \sim 0.1$, 
where ${\cal V} \simeq t_0^{3/2}-(t_1+t_2)^{3/2}/\sqrt{2}-(t_3+t_4)^{3/2}/\sqrt{2}$.
The decay constants of $f_1$ and $f_2$ become super-Planckian for $n_1 \gtrsim n_2 ={\cal O}(10)$, where
$f_1 = (2n_1/\pi |\Delta_1|)f$ and $f_2 = (2n_2/\pi \Delta_2)f$. 
On the other hand, 
if a non-perturbative effect is generated by a pure $SU(8n_1)$ gauge theory with a holomorphic gauge coupling $8T_2 + 9T_4$
instead of the sixth term in the superpotential, i.e.,
\bea
W \supset A e^{-\frac{2\pi}{n_1}T_2} \to A e^{-\frac{2\pi}{8n_1}(8 T_2 + 9T_4)},
\eea
$\Delta_1 = 1/8$ and $a = \pi/n_1$ are obtained while $\Delta_2$ and $b$ are unchanged.
One then realizes the mild hierarchy $f_2 = {\cal O}(10^{-2})  f_1$ for $n_1 ={\cal O}(10)$ and 
$n_2 ={\cal O}(1)$ to generate the large running spectral index.

The SUSY particles acquire a heavy soft mass, $m_{\rm soft} \sim W_0/\log(1/W_0)$, 
comparable the gravitino mass $m_{3/2} \sim W_0$ or slightly lighter. Thus they
are out of the reach of the accelerators, but their mass scale affects the lightest 
Higgs mass~\cite{Giudice:2011cg,Hebecker:2012qp,Ibanez:2012zg,Hebecker:2013lha}
through radiative corrections. 

\section{Spectral index and tensor-to-scalar ratio}
It is known that multi-natural inflation can accommodate a rather large running of the spectral index~\cite{Czerny:2014wua,Kobayashi:2010pz}, which can relax the tension between BICEP2 and Planck. Interestingly, the required small modulations to the inflaton potential
is a built-in feature of multi-natural inflation. 

Here let us focus on another aspect of multi-natural inflation. The existence of multiple sinusoidal functions also
enables the predicted values of the spectral index and the tensor-to-scalar ratio to be widely deviated from those
for natural inflation or other monomial chaotic inflation models.\footnote{This is also possible in
 the polynomial chaotic inflation~\cite{Nakayama:2013jka,Nakayama:2013nya}.} Although it is possible to consider 
 such deviations together with a large running spectral index in the inflaton potential with three sinusoidal functions,
 we focus on the case without running to simplify our analysis.

\begin{figure}[t]
\begin{center}
\includegraphics[scale=0.35]{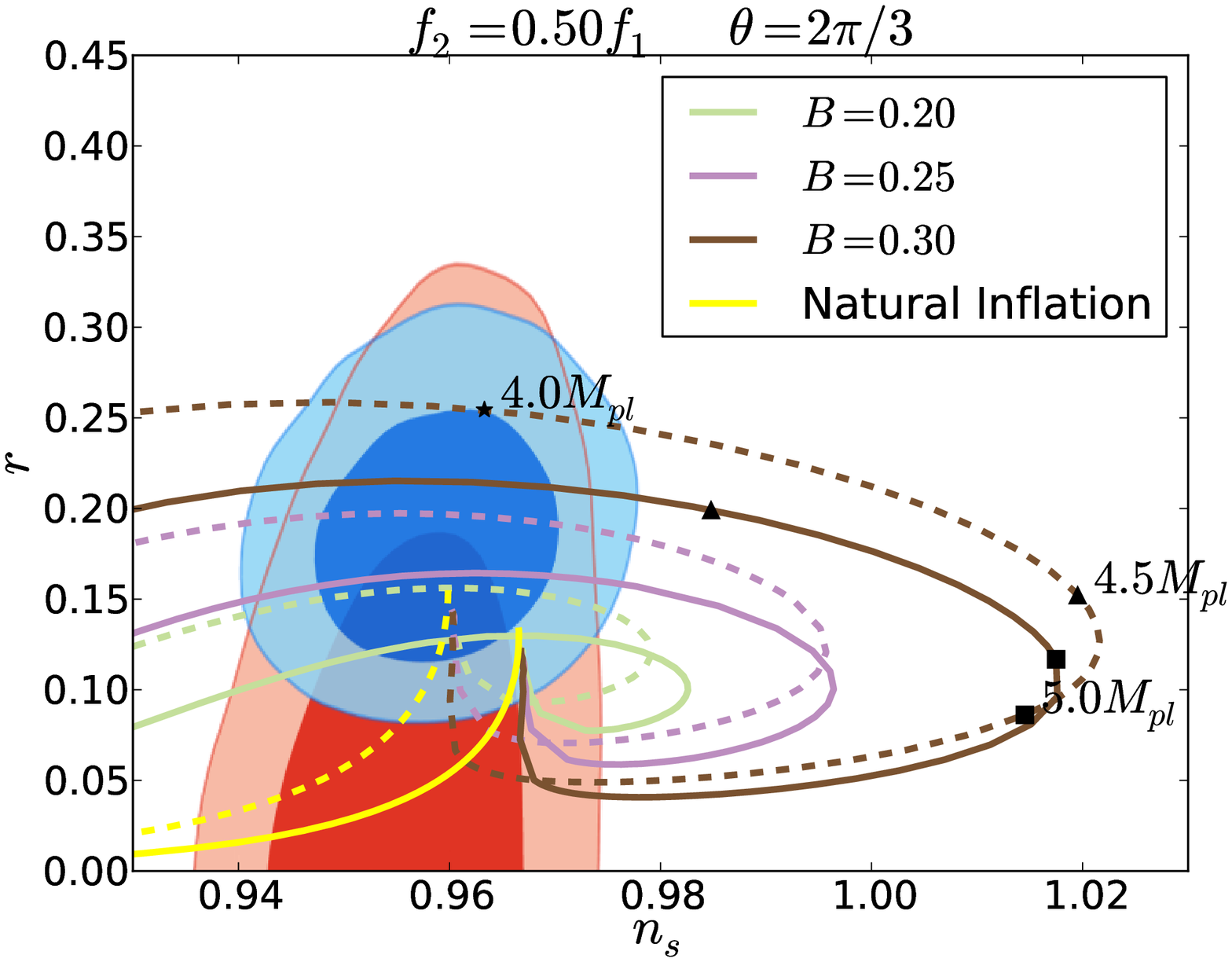}
\includegraphics[scale=0.345]{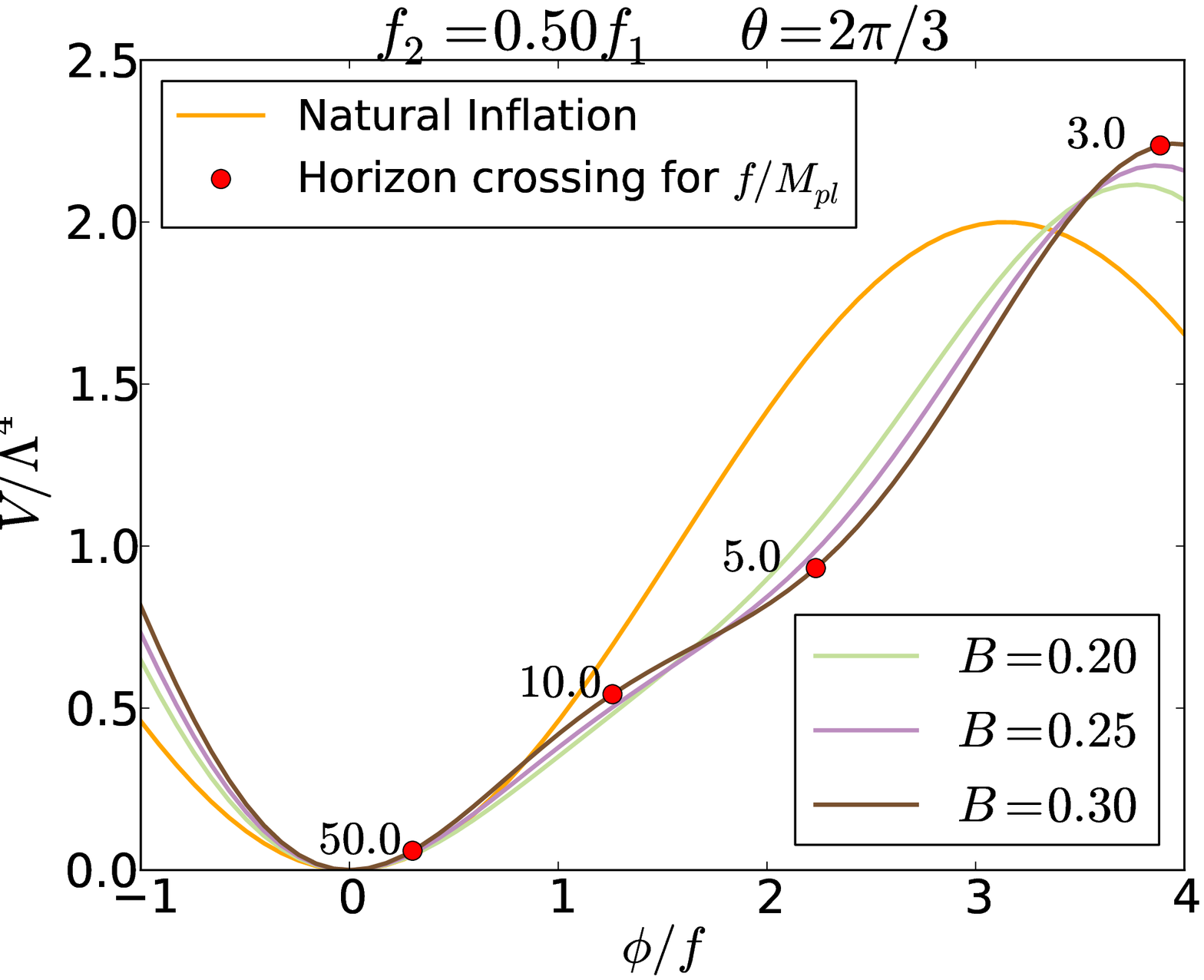}
\caption { 
Left: the prediction of $(n_s, r)$ of multi-natural inflation Solid (dashed) lines correspond to $N = 60$ ($N = 50$) e-folds before the
end of inflation. Predictions of $(n_s, r)$ for a few values of $f$ are shown on the brown dashed curve.
See the text for the explanation of the shaded regions. The allowed region will be shifted to larger values of $n_s$ if
the effective neutrino species is varied instead of the running spectral index. 
Right: the inflaton potential with multiple values for the decay constant $f$. The potential minimum is shifted to the origin for visualization purpose. When calculating the field excursion, we assume $N = 60$ e-folds.
}
\label{fig:nsr}
\end{center}
\end{figure}

We consider the multi-natural inflation with two sinusoidal functions with
\bea
  V(\phi) &=& C - \Lambda_1^4\cos(\f/f_1) - \Lambda_2^4\cos(\f/f_2 + \theta),
  \label{eq:eni}
\eea
where we set
\bea
f_2 = 0.5 f_1,~~\Lambda_2^4 = B \Lambda_1^4,~~ \theta = \frac{2 \pi}{3},
\eea
and the constant $C$ is fixed so that the inflaton potential at the potential minimum vanishes. 
We numerically solved the inflation dynamics for three case of $B = 0.20, 0.25$, and $0.30$ by
varying $f$. The predicted values of $(n_s, r)$ as well as the inflaton potential 
are shown in Fig.~\ref{fig:nsr}. The solid (dashed) lines correspond to $N = 60$ ($N = 50$) e-folds before the
end of inflation. Note that, while the lower shaded (red) region shows the region preferred by
{\it Planck}+WP+highL data allowing the running spectral index to vary, 
 the upper shaded (blue) region is obtained by combining the BICEP2. When the effective neutrino species $N_{\rm eff}$
is varied instead of the running, the allowed region is expected to be shifted to larger values of $n_s$ around unity
with $r = 0.1 \sim 0.2$. 
In the right panel, we assume $N = 60$. We also denote several predicted values of $(n_s, r)$ for different
values of the decay constant $f$ as a reference.
Interestingly, we can see from the figure that the spectral index close or above
unity together with $r = 0.1 \sim  0.2$ can be realized by multi-natural inflation.
Such features will be favored if the tension between BICEP2 and 
 Planck is solved by other sources such as dark radiation,  HDM,  or non-zero neutrino mass~\cite{Giusarma:2014zza}.

\section{Discussion and Conclusions}
\label{Secsum}

We have revisited multi-natural inflation~\cite{Czerny:2014wza} in supergravity for a UV completion.
In this model the inflaton potential  consists of two or more sinusoidal functions with different height and
 periodicity.
A further UV completion based on a string-inspired framework has been also considered.
The moduli stabilization of extra dimensions has to be done to give a sensible string theory in four dimensions.
Multi-natural inflation may be realized as a remnant of such moduli stabilization based on
various non-perturbative effects.

From an observational point of view, multi-natural inflation is interesting because
the existence of small modulations to the inflaton potential, which are necessary to generate a
large running spectral index, is a built-in feature. The running spectral index is one way to relax the tension 
between BICEP2 and Planck. 
For this to work, we need a mild hierarchy between the two decay constants. On the other hand, the tension 
could be solved by other sources such as dark radiation\footnote{
The existence of dark radiation may be ubiquitous in the Large Volume Scenario~\cite{Cicoli:2012aq,Higaki:2013lra,Angus:2014bia}.
}, 
in which case the scalar spectral index close to or even above unity is favored. We have shown that
the multi-natural inflation with two sinusoidal functions can realize a wide range of values of $(n_s, r)$
that include such cases.

\section*{Acknowledgment}
We thank Takeshi Kobayashi for useful discussion. 
This work was supported by Grant-in-Aid for  Scientific Research on Innovative
Areas (No.24111702, No. 21111006, and No.23104008) [FT], Scientific Research (A)
(No. 22244030 and No.21244033) [FT], and JSPS Grant-in-Aid for Young Scientists (B)
(No. 24740135 [FT] and No. 25800169 [TH]), and Inoue Foundation for Science.
This work was also supported by World Premier International Center Initiative
(WPI Program), MEXT, Japan [FT].

\appendix

\end{document}